\documentclass{article}
\usepackage{ijcai21}

\usepackage{times}
\usepackage{soul}
\usepackage{url}
\usepackage[hidelinks]{hyperref}
\usepackage[utf8]{inputenc}
\usepackage[small]{caption}
\usepackage{graphicx}
\usepackage{amsmath}
\usepackage{amsthm}
\usepackage{amssymb}
\usepackage{multirow}
\usepackage{subfigure}
\usepackage{booktabs}
\usepackage{algorithm}
\usepackage{xcolor}
\usepackage{algorithmic}
\usepackage{diagbox}
\usepackage{booktabs}
\usepackage{comment}
\newcommand{\tabincell}[2]{
\begin{tabular}{@{}#1@{}}#2\end{tabular}
}
\newtheorem{example}{Example}

\title{Regulating Ownership Verification for Deep Neural Networks: \\ Scenarios, Protocols, and Prospects}
\author{
Fang-Qi Li$^1$\and
Shi-Lin Wang$^1$\footnote{Shi-Lin Wang is the contact author.}\And
Alan Wee-Chung Liew$^2$
\affiliations
$^1$School of Electronic Information and Electrical Engineering, \\Shanghai Jiao Tong University, Shanghai, China.\\
$^2$School of Information and Communication
Technology, \\Griffith University, Gold Coast Campus, Australia.
\emails
\{solour\_lfq, wsl\}@sjtu.edu.cn, a.liew@griffith.edu.au
}

\begin{document}

\maketitle

\begin{abstract}
With the broad application of deep neural networks, the necessity of protecting them as intellectual properties has become evident.
Numerous watermarking schemes have been proposed to identify the owner of a deep neural network and verify the ownership.
However, most of them focused on the watermark embedding rather than the protocol for provable verification.
To bridge the gap between those proposals and real-world demands, we study the deep learning model intellectual property protection in three scenarios: the ownership proof, the federated learning, and the intellectual property transfer.
We present three protocols respectively.
These protocols raise several new requirements for the bottom-level watermarking schemes.
\end{abstract}

\section{Introduction}
The development of deep learning boosted the application of deep neural networks (DNN).
Given abundant data and computing resources, DNNs outperform traditional models in many disciplines such as image processing, natural language processing~\cite{guo2020gluoncv}, internet of things~\cite{lv2020deep}, etc.
The expense behind a DNN is high.
Much data is collected, processed, and labeled.
Designing the DNN architecture and tuning its parameters also involves tremendous effort.
Therefore, DNNs are the intellectual property (IP) of the legitimate owner.

Ownership verification (OV) is necessary for the Intellectual Property Protection (IPR) of DNNs.
To achieve OV, various DNN watermarking schemes have been proposed.
A watermarking scheme embeds the owner-dependent watermark into the DNN, whose later revealing proves the owner's identity.
Based on the access level at which the suspicious DNN can be interacted with, watermarking schemes can be classified into \textit{white-box} ones and \textit{black-box} ones.

In the white-box setting, the owner has full access to the pirated DNN.
The watermark can be encoded into the model's parameters~\cite{uchida2017embedding} or intermediate outputs~\cite{darvish2019deepsigns}.
The owner can also insert extra modules into the DNN's intermediate layers for OV~\cite{fan2021deepip}.
As for the black-box setting, the owner can only interact with the pirated model through an API.
Watermarking schemes for this case usually resort to backdoors~\cite{zhang2018protecting,zhu2020secure}.

In contrast to the variety of proposed watermarking schemes, discussions on the protocol under which the OV is conducted remain scanty.
The OV protocol is indispensable for commercializing deep learning models, 
but established works on OV protocols mainly focused on the secure transmission of watermarks and are highly inflexible~\cite{adi2018turning}.
So the IPR of DNNs as a service remains a challenge.

Emerging real-world scenarios are providing diversified challenges for the verification protocols.
For example, in the model competition, there exists a trusted sponsor with white-box access to all DNNs.
While in commercial services where models are deployed as APIs, such a trusted party is unavailable.
Distributed learning paradigms such as the Federated Learning (FL) introduce extra security requirements, which go beyond the scope of established watermarking schemes and protocols.
Moreover, it is of broad interest to enlarge the coverage of IPR for DNN models to disciplines other than piracy identifying, e.g., secure intellectual property transfer.

To apply established DNN watermarking schemes to real-world scenarios for IPR of DNN models, we formulate these practical demands and propose candidate protocols meeting respective settings.
The contributions of this paper are:
\begin{itemize}
\item We analyze three real-world scenarios involving IPR of DNNs and formulate respective protocols.
\item We show that some security properties of the proposed protocols can be built upon the security of underlying watermarking schemes by reduction.
\item We explore several additional requirements for the watermarking schemes introduced by the protocols, which are prospective directions for further research.
\end{itemize}

\section{Properties of Watermarking Schemes}
\label{section:2}
In general, a watermarking scheme $\texttt{WM}$ is composed of two modules $\left\{\texttt{Gen},\texttt{Embed} \right\}$, one generates the watermark:
$$\texttt{key}\leftarrow\texttt{Gen}(1^{N}),$$
one embeds the watermark into the model to be protected:
$$(M_{\text{WM}},\texttt{verify})\leftarrow\texttt{Embed}(M_{\text{clean}},\texttt{key}).$$
The watermark is an identifier $\texttt{key}$ representing the owner's identity and $N$ is the security parameter.
The embedding module takes a clean DNN $M_{\text{clean}}$ as its input.
The module $\texttt{verify}$ returned from $\texttt{Embed}$ is part of the evidence for reconstructing the owner's identity from $M_{\text{WM}}$ to achieve OV.
As has been outlined in~\cite{ours}, a watermarking scheme has to satisfy the following security requirements.

\subsection{Correctness}
The module $\texttt{verify}$ can identify the owner's identity from the watermarked model:
$$\text{Pr}\left\{\texttt{verify}(M_{\text{WM}},\texttt{key})=1 \right\}\geq1-\epsilon,$$
where $\epsilon$ is a function negligible in $N$.
Meanwhile, an adversary's identity cannot pass the verifier:
$$\text{Pr}\left\{\texttt{verify}(M_{\text{WM}},\texttt{key}_{\text{ADV}})=0 \right\}\geq1-\epsilon,$$
where $\texttt{key}_{\text{ADV}}$ is the adversary's evidence randomly sampled from the key space.

\subsection{Robustness}
The adversary can tune the pirated model using fine-tuning, neuron-pruning, fine-pruning~\cite{liu2018fine}, even distillation~\cite{zhang2021deep}:
$$M_{\text{tuned}}\xleftarrow{\text{tuning}}M_{\text{WM}}.$$
Under a robust watermarking scheme, such tuning should not affect the accuracy of OV:
$$\text{Pr}\left\{\texttt{verify}(M_{\text{tuned}},\texttt{key})=1 \right\}\geq1-\epsilon.$$

\subsection{Covertness}
\label{section:covertness}
An adversary should not be able to distinguish a watermarked model from a clean one.
Otherwise, the adversary might manage to escape the IP regulation.
Formally, we design the following Algo.~\ref{exp:1}.
\begin{algorithm}[htbp]
\caption{$\texttt{Exp}^{\text{covertness}}_{\mathcal{\mathcal{A}}}$.}
\label{exp:1}
\textbf{Input}: $\mathcal{A}$, $N$, $\texttt{WM}$, $M_{\text{clean}}$\\
\textbf{Output}: Whether $\mathcal{A}$ wins or not
\begin{algorithmic}[1]
\STATE Randomly select $b\leftarrow \left\{0,1 \right\}$.
\STATE Generate $M_{\text{WM}}$ from $\texttt{WM}(M_{\text{clean}},N)$.
\STATE $\mathcal{A}$ is given $N$ and $\texttt{WM}$. 
\IF {$b=0$}
\STATE $\mathcal{A}$ is given $M_{\text{clean}}$.
\ELSE
\STATE $\mathcal{A}$ is given $M_{\text{WM}}$.
\ENDIF
\STATE $\mathcal{A}$ outputs $\hat{b}$.
\STATE $\mathcal{A}$ wins the experiment if $\hat{b}=b$.
\end{algorithmic}
\end{algorithm}
The watermarking scheme is covert if no efficient probabilistic machine $\mathcal{A}$ can win $\texttt{Exp}^{\text{covertness}}_{\mathcal{\mathcal{A}}}$ with a probability significantly higher than $\frac{1}{2}$.

\subsection{Privacy-preserving}
The privacy-preserving property suggests that no adversary can identify the model's ownership given only partial information of the owner.
One type of privacy-preserving is defined through Algo.~\ref{exp:2}.
\begin{algorithm}[htbp]
\caption{$\texttt{Exp}^{\text{key-pp}}_{\mathcal{\mathcal{A}}}$.}
\label{exp:2}
\textbf{Input}: $\mathcal{A}$, $N$, $\texttt{WM}$, $\texttt{key}_{0}\neq\texttt{key}_{1}$, $M_{\text{clean}}$\\
\textbf{Output}: Whether $\mathcal{A}$ wins or not
\begin{algorithmic}[1]
\STATE Randomly select $b\leftarrow \left\{0,1 \right\}$.
\STATE Generate $M_{\text{WM}}$ from $\texttt{WM}(M_{\text{clean}},\texttt{key}_{b},N)$.
\STATE $\mathcal{A}$ is given $M_{\text{WM}}$, $M_{\text{clean}}$, $N$, $\texttt{WM}$, $\texttt{key}_{0}$, $\texttt{key}_{1}$.
\STATE $\mathcal{A}$ outputs $\hat{b}$.
\STATE $\mathcal{A}$ wins the experiment if $\hat{b}=b$.
\end{algorithmic}
\end{algorithm}
If no efficient $\mathcal{A}$ can win $\texttt{Exp}^{\text{key-pp}}_{\mathcal{\mathcal{A}}}$ with a probability significantly higher than $\frac{1}{2}$ then $\texttt{WM}$ is \textit{key-privacy-preserving}~\cite{ours}.
Analogously, we can define \textit{verifier-privacy-preserving}.

The key-privacy-preserving properties suggests that the $\texttt{verify}$ module of the watermarking scheme should depend on $\texttt{key}$. 
Otherwise the privacy is easily breached. 

\begin{example}
The watermarking scheme of Uchida's replaces $U$ parameters within the clean DNN by special digits.
Its keyspace can be defined as $\mathbb{R}^{U}$ or $\Theta^{U}\times\mathbb{R}^{U}$, where $\Theta$ is the space of all parameters in the DNN model.
Being formulated in the first manner, Uchida's is a key-privacy-preserving scheme.
In the second formulation, a legal $\texttt{key}$ includes both the place where the digits are embedded and the digits.
The corresponding $\texttt{verify}$ is only a parameter-free comparison operator so it is not a key-privacy-preserving scheme.
\end{example}

\subsection{Overwriting issues}
Having known the watermarking scheme, the adversary can embed its identity into the model:
$$(M_{\text{OW}},\texttt{verify}_{\text{OW}})\leftarrow\texttt{Embed}(M_{\text{WM}},\texttt{key}_{\text{ADV}}).$$
So the ownership becomes ambiguous.
To cope with this threat, the owner's watermark must not be invalidated, i.e.:
$$\text{Pr}\left\{\texttt{verify}(M_{\text{OW}},\texttt{key})=1 \right\}\geq1-\epsilon.$$
In cases where the adversary embeds its watermark into the DNN and \textit{redeclare} the ownership, extra mechanisms, e.g., authorized time-stamp, are necessary to break the tie.

\section{Scenarios and Watermarking Protocols}
\label{section:3}
IPR involves proving the ownership to a third-party, which we denote as the \textit{notary}.
Embedding and recovering watermarks without clarifying the role of the notary is insufficient for IPR.
Top-level protocols, follow which all parties involved in IPR (the owner, the adversary, and the notary) operate, are indispensable.
The configuration of the three parties' functionalities varies in different scenarios, so it is necessary to design a specialized protocol for each case.
We present practical protocols for three important real-world scenarios:
\begin{itemize}
\item An owner proves its ownership over a DNN to a notary.
\item Collaborated owners in FL prove their ownership over a DNN to a notary, during which they can recover each other's identity proof and trace potential traitors.
\item An owner transfers the IP of its DNN to a third party.
\end{itemize}
For watermarking schemes, these protocols introduce extra security requirements besides those listed in Section~\ref{section:2}.

\subsection{Protocols for ownership proof}
\label{section:3.1}
\subsubsection{The centralized OV protocol}
The simplest OV protocol is centralized, in which the notary is a verification center responsible for publishing legitimate ownership proofs.
This is the case which most established watermarking schemes have assumed.
It involves two steps:
\begin{enumerate}
\item The owner submits $(\texttt{key},\texttt{verify})$ and the access to $M$ to the notary.
\item The notary computes $\texttt{verify}(M,\texttt{key})$ and publishes the output.
\end{enumerate}
To use white/black-box watermarking schemes, the owner has to provide the white/black-box access of the suspicious model to the notary.
To preserve privacy, the channel between the owner and the notary has to be encrypted.
As for a curious notary, a Secure Multi-Party Computation (SMPC)~\cite{bogetoft2009secure} protocol can be adopted to protect the owner's data.
Using such a protocol, the redeclaration problem can be solved.
Instead of generating $\texttt{key}$ on its own, the owner queries the notary for time authorization, who would return a key containing the time-stamp back to the owner.
Overwriting and redeclaring cannot falsify the time-stamp so the ownership is secured.

Despite its simplicity, this protocol has many defects:
\begin{itemize}
\item The proof is valid only within the community that recognizes the notary's credit.
It is difficult to accommodate this protocol for a broader range of entities.
\item If the notary is compromised then all verifications within the community are at risk.
\item Attacks against centralized protocols, such as the Deny Of Service (DOS) can paralyze the protocol.
\end{itemize}

\subsubsection{The decentralized OV protocol}
Given the defects of the centralized protocol, we propose a decentralized protocol for OV~\cite{ours}.
Instead of relying on a verification center, we resort to a community of agents distributed across the network.
To prove its ownership over a DNN, the owner broadcasts the necessary evidence to the verification community.
Then each agent can volunteer to conduct the verification and broadcast the result.
The OV is finished by voting through the entire community.
To solve the redeclaration dilemma, an owner has to broadcast the hash of the DNN architecture and the evidence under a consensus protocol~\cite{ongaro2014search}.
Then the entire community would have a consensus on the time-stamp corresponding to ownership.
This protocol is outlined in Algo.~\ref{ptc:decentralized}.
Its unforgeability and correctness can be reduced to the security of $\texttt{WM}$, that of the digital signature scheme, and the reliability of the consensus protocol.

\begin{algorithm}[htbp]
\caption{The decentralized OV protocol.}
\label{ptc:decentralized}
\textbf{Participants}: The owner, the verification community\\
\textbf{Modules}: A watermarking scheme $\texttt{WM}$, a digital signature scheme, a consensus protocol
\begin{algorithmic}[1]
\STATE The owner generates $M_{\text{clean}}$.
\STATE The owner generates $\texttt{key}$, $M_{\text{WM}}$, and $\texttt{verify}$ by $\texttt{WM}$.
\STATE The owner signs the following message:
$$\langle\texttt{time}\|\texttt{hash}(\texttt{key})\|\texttt{hash}(\texttt{verify})\|\texttt{hash}(\text{info})\rangle$$
using the digital signature scheme, where $\texttt{time}$ is the current time-stamp, $\texttt{hash}$ is a hash function, and $\texttt{info}$ descripts the DNN model's architecture.
\STATE The owner broadcasts the signed message to the community using the consensus protocol.
\STATE To conduct OV over a DNN $M$, the owner signs and broadcasts $\langle M\|\texttt{key}\|\texttt{verify}\rangle$.
\STATE An agent retrives the time-stamp by computing $\texttt{hash}(\texttt{verify})$, and submits $\texttt{verify}(M,\texttt{key})$ to the community using the consensus protocol.
\end{algorithmic}
\end{algorithm}

\begin{figure*}[!htbp]
\centering
\subfigure[The first verification.]{
\begin{minipage}[htbp]{0.3\linewidth}
\centering
\includegraphics[width=5.5cm]{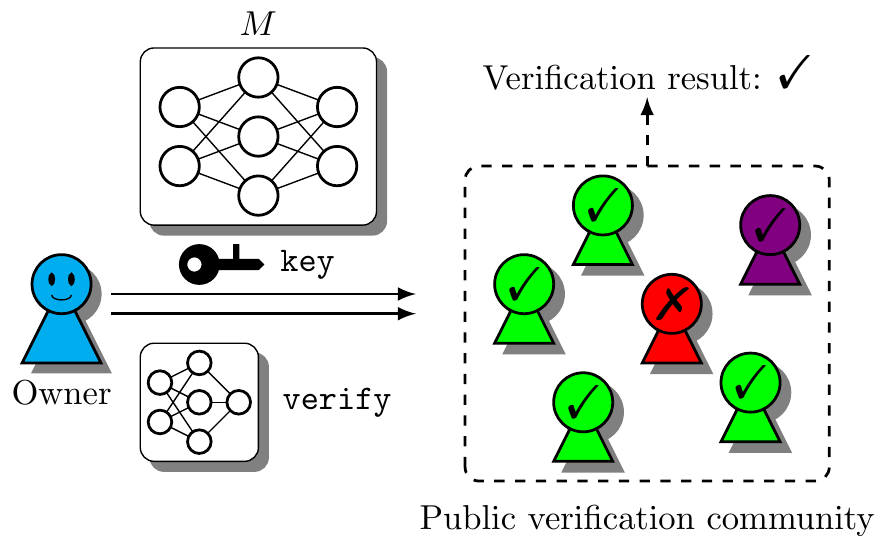}
\end{minipage}%
}%
\subfigure[The spoil attack.]{
\begin{minipage}[htbp]{0.3\linewidth}
\centering
\includegraphics[width=5cm]{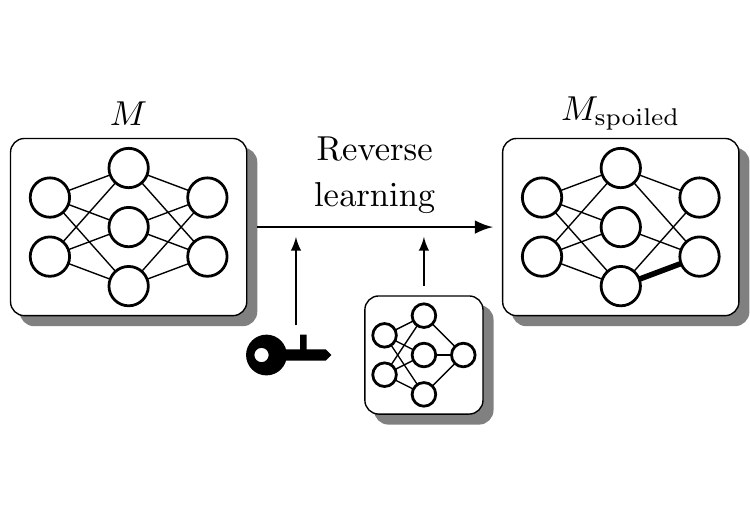}
\end{minipage}%
}%
\subfigure[The second verification.]{
\begin{minipage}[htbp]{0.3\linewidth}
\centering
\includegraphics[width=5.5cm]{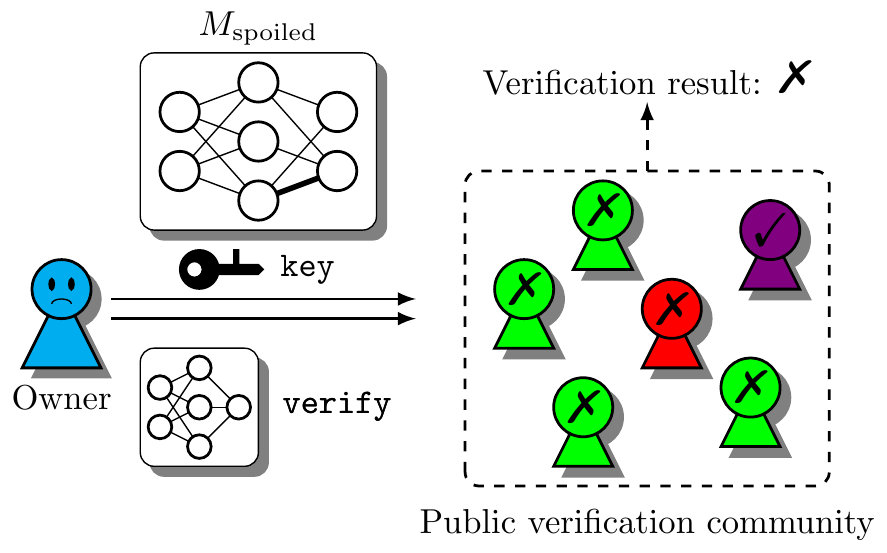}
\end{minipage}%
}%
\caption{The spoil attack. The blue node is the owner, green nodes are benign agents, the red node is a malicious agent, and the purple one is the eavesdropping adversary.}
\label{figure:spoil}
\end{figure*}

As in other distributed service systems~\cite{mengelkamp2018blockchain}, to motivate the entire community to conduct OV, each correct verification assigns credits to agents that contribute to the proof, with which they can initiate their OV requests.
This protocol is immune to attacks that only compromise a single agent.
However, the communication traffic is increased.
Especially when the owner adopts a white-box watermarking scheme, then each agent has to download the entire model.
Since the proof is done on many independent agents, using SMPC thoroughly would be expensive and inefficient.
Therefore, an eavesdropping adversary may steal the evidence corresponding to the owner and its model.
Then the adversary can \textit{spoil} this specific watermark so the owner can no longer succeed in OV over the new model, this \textit{spoil attack} is illustrated in Fig.~\ref{figure:spoil}.

\subsubsection{Discussion: the spoil attack}
\label{section:3.1.3}
The spoil attack, as an additional threat in the decentralized OV protocol, has seldom entered the concern of designers of DNN watermarking schemes.
Consequently, almost all established watermarking schemes are vulnerable against the spoil attack, which fact challenges the applicability of the decentralized OV protocol.

Backdoor-based watermarking schemes can be spoiled by fitting the DNN to randomly shuffled labels on the triggers.
For white-box watermarking schemes, the adversary can spoil the watermark by tuning the model reversely.

Theoretically, the security against the spoil attack can be defined through Algo.~\ref{exp:3}.
\begin{algorithm}[htbp]
\caption{$\texttt{Exp}^{\text{spoil}}_{\mathcal{\mathcal{A}},\delta}$.}
\label{exp:3}
\textbf{Input}: $\mathcal{A}$, $N$, $\texttt{WM}$, $M_{\text{clean}}$\\
\textbf{Output}: Whether $\mathcal{A}$ wins or not
\begin{algorithmic}[1]
\STATE Generate $M_{\text{WM}}$, $\texttt{key}$ and $\texttt{verify}$ from $\texttt{WM}(M_{\text{clean}},N)$.
\STATE $\mathcal{A}$ is given $M_{\text{WM}}$, $\texttt{key}$, $N$, $\texttt{WM}$ and $\texttt{verify}$.
\STATE $\mathcal{A}$ outputs $M_{\text{spoiled}}$.
\STATE $\mathcal{A}$ wins the experiment if $\texttt{verify}(M_{\text{spoiled}},\texttt{key})=0$ and $M_{\text{spoiled}}$'s performance declines for no larger than $\delta$ compared with $M_{\text{WM}}$.
\end{algorithmic}
\end{algorithm}
The scheme $\texttt{WM}$ is secure against the spoil attack iff no efficient adversary can win $\texttt{Exp}^{\text{spoil}}_{\mathcal{A},\delta}$ with non-negligible probability for a given $\delta$.

Such proof is intractable for almost all established watermarking schemes.
It is unknown whether a scheme provably secure against the spoil attack exists or not.

As a substitute, we can improve traditional watermarking schemes against the spoil attack by simply embedding multiple watermarks into the DNN to be protected.
Since each round of OV only exposes one watermark, such configuration can resist the spoil attack.
But inserting multiple watermarks also calls for additional requirements, namely the watermarking capacity and independence.

As defined in~\cite{li2021practical}, the $(\delta,\texttt{WM})$-\textit{watermarking capacity} for a DNN model $M$, $\texttt{cap}^{\delta}_{\texttt{WM}}$, is the maximum number of watermarks that can be embedded and verified correctly before $M$'s performance declines for $\delta$.
This OV service can survive $\texttt{cap}^{\delta}_{\texttt{WM}}$ rounds of spoil attacks by sacrificing the performance for at most $\delta$.

The other aspect is: spoiling one watermark should not affect other ones.
Otherwise, spoiling one watermark might invalidate others that have not been exposed and decrease the times of correct OV.
To evaluate the \textit{watermarking independence} against the spoil attack of $\texttt{WM}$ w.r.t. a DNN $M$, we firstly insert $Q$ watermarks into $M$ using $\texttt{WM}$.
Then we spoil a random watermark and denote the number of watermarks that can still be correctly verified as $r$.
The higher the \textit{watermarking independence score} $\frac{r}{Q}$ is, the more robust $\texttt{WM}$ is against the spoil attack.

\subsection{The OV protocol for federated learning}
\begin{figure}[htbp]
\centering
\includegraphics[width=7cm]{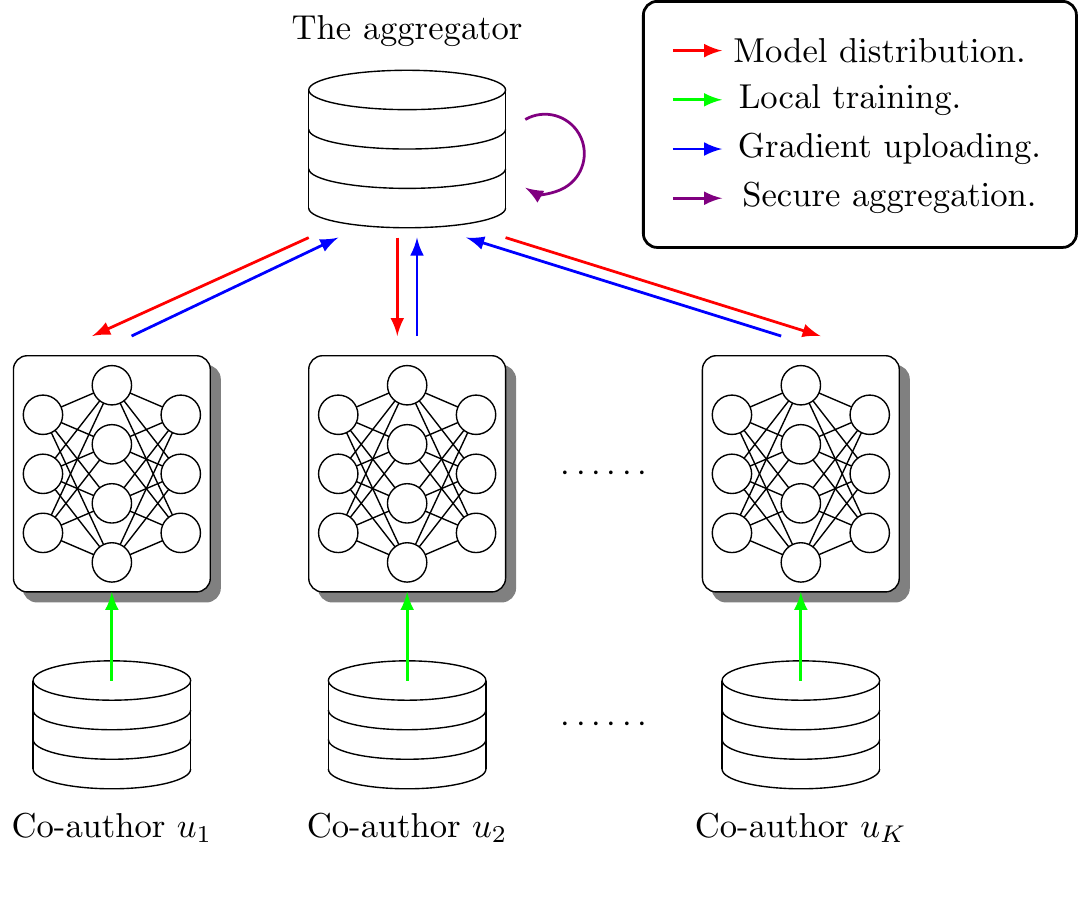}
\caption{The client-server architecture for FL.}
\label{figure:fl}
\end{figure}

In the basic OV protocol, each DNN has a unique owner.
The development of distributed learning paradigms, especially FL~\cite{yang2019federated}, has changed this assumption.
In FL, many parties coordinated by an aggregator cooperate to train one deep learning model without interchanging local data as illustrated in Fig.~\ref{figure:fl}.
Each participating party should be able to verify its ownership over the model independently.
The privacy of each owner must not be breached, concretely, an owner can not falsify himself as another owner.
Considering the collaboration of all owners, it is desirable that when one owner undergoes severe spoil attacks so its identity information is erased from the model, other owners can help it recover the ownership proof.
Moreover, if one party betrays its co-authors, pirates the intermediate model and claims it to be its product, then the honest parties can correctly identify this traitor.

These four requirements, \textit{independent verification}, \textit{privacy-preserving}, \textit{recovery}, and \textit{traitor-tracing}, mark the characteristics for OV in FL.
To reduce the communication traffic between the owners and the verification community, achieve the recovery property, and ensure traitor-tracing, a modified version of the basic decentralized OV protocol, $\texttt{Merkle-Sign}$, has been proposed~\cite{li2021practical}.
Its representative features are:
\begin{itemize}
\item As shown in Fig.~\ref{figure:MS1}, in training, the aggregator embeds its key ($\texttt{key}_{0}$), a surveillance key ($\texttt{key}_{k}^{\dagger}$) into the intermediate model distributed to the $k$-th author to achieve traitor-tracing.
When training terminates, the aggregator embeds the identity information of all authors into the model and broadcasts the hashed message as in the decentralized OV protocol.

\begin{figure}[htbp]
\centering
\includegraphics[width=8cm]{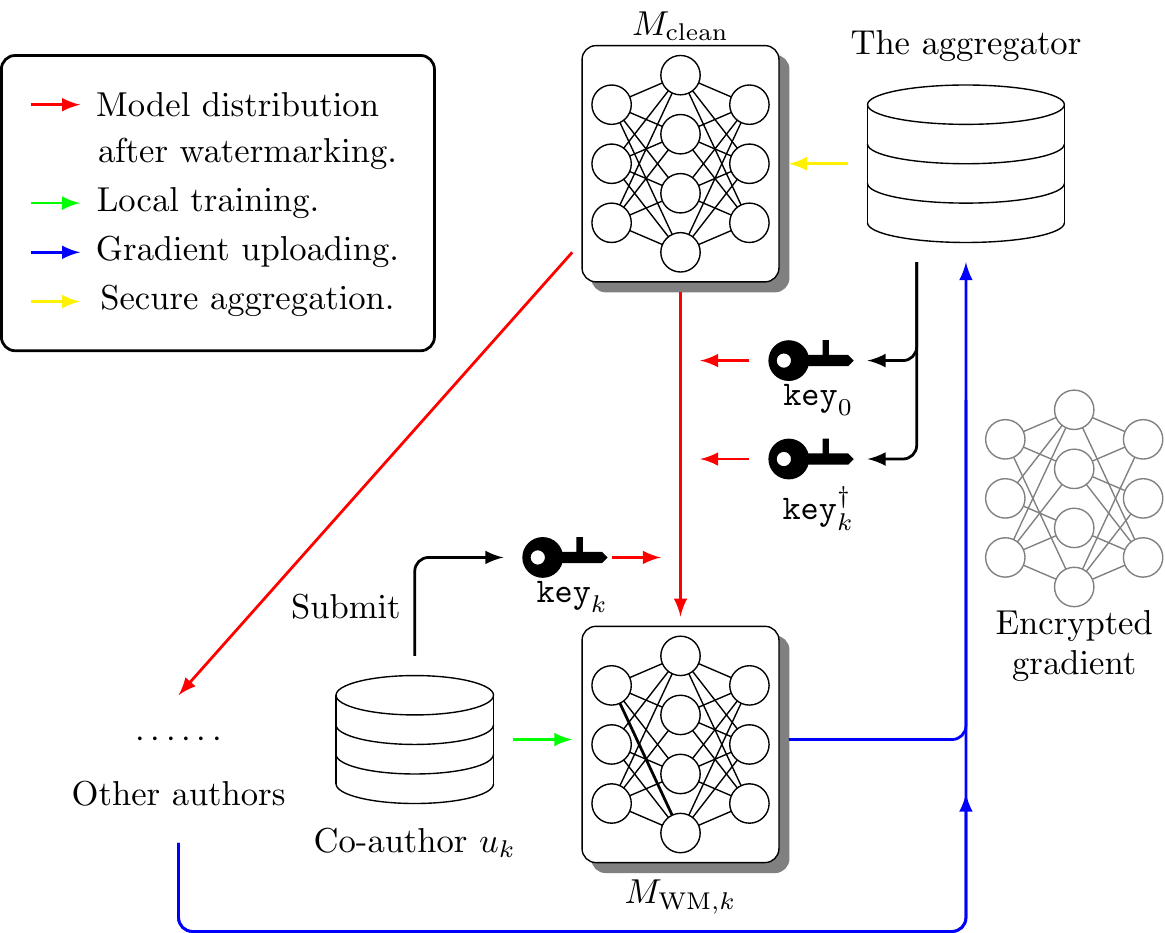}
\caption{The $\texttt{Merkle-Sign}$ watermarking framework for FL.}
\label{figure:MS1}
\end{figure}

\item When broadcasting messages to the verification community, the hashed value of the identity proof for all owners are associated into a Merkle-tree~\cite{li2013efficient} so owners can build correlations between their evidence to enable recovery.
\end{itemize}

The analysis in~\cite{li2021practical} showed that the security of this protocol and four characteristics can be reduced to the computational hardness of the cryptological primitives within.

The watermarking scheme for $\texttt{Merkle-Sign}$ also has to have a large watermarking capacity and independence.
In addition, the embedding process is expected to be efficient and exert only slight modification to the entire DNN model.
Otherwise, the model might fail to converge.

\subsubsection{Discussion: aggregatable watermarks}
In $\texttt{Merkle-Sign}$, the aggregator is in charge of embedding watermarks into the DNN model.
Therefore this protocol needs a trusted aggregator and is not completely decentralized as in recent configurations of secure FL~\cite{wei2020federated}.

To transfer the responsibility of watermark embedding from the aggregator to the owners, it is necessary that: the verification process remains valid to the aggregator's model aggregation.
We denote the aggregator's combinator as $\mathcal{O}$, it can be model average, ensembling, etc.
For $K$ independent owners, this requirement can be formulated as:
\begin{equation}
\begin{aligned}
&\forall k, \texttt{key}_{k}\leftarrow\texttt{Gen}(1^{N}),M_{\text{WM},k}\leftarrow M_{\text{clean},k},\\
&\ \ \ \ \ \ (M_{\text{WM},k},\texttt{verify}_{k})\leftarrow\texttt{Embed}(M_{\text{WM},k},\texttt{key}_{k}),\\
&M_{\text{WM}}\leftarrow \mathcal{O}(M_{\text{WM},1}\cdots M_{\text{WM},K}),\\
&\forall k, \text{Pr}\left\{\texttt{verify}_{k}(M,\texttt{key}_{k})1 \right\}\geq 1-\epsilon.
\end{aligned}
\label{equation:aggregatable}
\end{equation}
In which $M_{\text{clean},k}$ is the model distributed to the $k$-th author from the aggregator in each epoch.
If the watermarking scheme $\texttt{WM}$ satisfies~\eqref{equation:aggregatable} then we define it as an \textit{aggregatable} scheme.
An aggregatable watermarking scheme can improve $\texttt{Merkle-Sign}$ into a completely decentralized protocol regarding both model training and OV.

\subsection{The Protocol for DNN IP transfer}
\label{section:3.3}
Apart from OV, IPR in DNN commercialization includes many other aspects, among which an important one is the transferring of DNN as IP. 
For the purchaser of a DNN, it is important that the deep learning model he paid for only contains his identity information. 
Otherwise, the seller might unilaterally cancel the transaction by redeclaring the ownership over this DNN from the seller's watermarks hidden within. 
Therefore, it is necessary to convince the purchaser that the sold DNN product is free from any watermark. 
Concretely, such convincing is possible only if there exists an algorithm $\mathcal{P}$ that can win the experiment defined in Algo.~\ref{exp:transfer} with probability one.
\begin{algorithm}[htbp]
\caption{$\texttt{Exp}^{\text{clean}}_{\mathcal{\mathcal{P}}}$.}
\label{exp:transfer}
\textbf{Input}: $\mathcal{P}$, $N$, $\texttt{WM}$, $M_{\text{clean}}$\\
\textbf{Output}: Whether $\mathcal{P}$ wins or not
\begin{algorithmic}[1]
\STATE Generate $M_{\text{WM}}$ from $\texttt{WM}(M_{\text{clean}},N)$.
\STATE Randomly select $b\leftarrow \left\{0,1 \right\}$.
\STATE $\mathcal{P}$ is given $N$ and $\texttt{WM}$. 
\IF {$b=0$}
\STATE $\mathcal{P}$ is given $M_{\text{clean}}$.
\ELSE
\STATE $\mathcal{P}$ is given $M_{\text{WM}}$.
\ENDIF
\STATE $\mathcal{P}$ outputs $\hat{b}$.
\STATE $\mathcal{P}$ wins the experiment if $\hat{b}=b$.
\end{algorithmic}
\end{algorithm}

Notice that in $\texttt{Exp}^{\text{clean}}_{\mathcal{\mathcal{P}}}$, $\mathcal{P}$ is given neither $\texttt{key}$ nor $\texttt{verify}$ since the seller might hide them from the purchaser. 
We assumed that the watermarking scheme and the security parameter have been agreed on within the community where the transaction takes place. 
To conduct a DNN IP transfer the purchaser runs $\mathcal{P}$ on the model transmitted by the seller. 
If the output is zero then the purchaser is convinced that the sold model is free from any watermark. 
Then the purchaser can treat this model as its $M_{\text{clean}}$, deploy services by it, and protect it as his IP by protocols in Section~\ref{section:3.1}. 

It is remarkable that the existence of a distinguisher $\mathcal{P}$ winning $\texttt{Exp}^{\text{clean}}_{\mathcal{\mathcal{P}}}$ is contradictive to the fundamental property of covertness defined in Section~\ref{section:covertness}. 
So a watermarking scheme designed for one purpose is not necessarily a option in another scenario. 


\section{Experiments and Discussions}
The evaluation of watermarking schemes w.r.t. basic security requirements has been presented in~\cite{chen2018performance}.
To examine the adaptivity of current watermarking schemes to the real-world settings and corresponding protocols, we are interested in additional requirements listed in Table~\ref{table:1}.

\begin{table*}[htbp]
\centering
\begin{tabular}{c|c|c|c|c|c}
\toprule
\diagbox [width=8em,trim=l]{Protocol}{Metric} & \tabincell{c}{Performance\\decline due to\\the spoil attack \\ \textbf{(A)}} & \tabincell{c}{Watermark \\ capacity\\ \ \\ \textbf{(B)}} & \tabincell{c}{Watermark\\ independence\\ score\\ \textbf{(C)}} & \tabincell{c}{Time consumption \\ of watermark\\ embedding\\ \textbf{(D)}} & \tabincell{c}{Performance \\ decline in FL due\\ to watermarking \\ \textbf{(E)}} \\
\midrule
Decentralized OV & \checkmark & \checkmark & \checkmark & -- & --  \\
\tabincell{c}{$\texttt{Merkle-Sign}$} & \checkmark & \checkmark & \checkmark & \checkmark & \checkmark \\
DNN IP transferring & \checkmark & $\times$ & \checkmark & -- & --  \\
\bottomrule
\end{tabular}
\caption{Additional security requirements.
\checkmark denotes necessity, -- denotes irrelevance, and $\times$ denotes negativity.}
\label{table:1}
\end{table*}

\subsection{Settings}
We adopted ResNet-50~\cite{he2016deep} as the backbone DNN architecture.
Experiments were conducted on three datasets: MNIST~\cite{deng2012mnist}, CIFAR10, and CIFAR100~\cite{krizhevsky2009learning}.
To evaluate the adaptivity of established DNN watermarking schemes to the presented protocols, we considered five candidates: Uchida's, random trigger ($\texttt{Rand}$), $\texttt{Wonder Filter}$ ($\texttt{WF}$), $\texttt{ATGF}$, and $\texttt{MTL-Sign}$ ($\texttt{M-S}$).
In Uchida's, we adopted $U=20$.
For random trigger and $\texttt{WF}$, we adopted the configuration in~\cite{zhang2018protecting} and~\cite{li2019persistent}.
As for $\texttt{ATGF}$ and $\texttt{MTL-Sign}$, the initialization in~\cite{li2021practical} and~\cite{ours} were used.
All experiments were conducted under the $\texttt{PyTorch}$ framework.

\subsection{Evaluations of extra security requirements}
The metric \textbf{(A)} reflects the damage of the spoil attack to the DNN model.
The higher \textbf{(A)} is, the less likely an adversary is willing to conduct a spoil attack.
Metrics \textbf{(B)}, \textbf{(C)}, \textbf{(D)}, and \textbf{(E)} have been introduced in Section~\ref{section:3}.
We conducted spoil attacks against five watermarking schemes as described in Section~\ref{section:3.1.3}. 
To compute \textbf{(B)}, we adopted $\delta$ as the error rate of classification of the clean model. 
To compute \textbf{(C)}, we adopted $Q=50$.
To compute \textbf{(E)}, we included 200 independent authors in FL, and the aggregator used the model average for DNN model combination.
The evaluations of \textbf{(A)} to \textbf{(E)} in all datasets are presented in Table~\ref{table:m},~\ref{table:c10}, and~\ref{table:c100}.
The optimal scheme w.r.t. each metric is highlighted.

\begin{table}[htbp]
\centering
\begin{tabular}{m{1.6cm}<{\centering}|m{0.8cm}<{\centering}|m{1.1cm}<{\centering}|m{0.8cm}<{\centering}|m{0.8cm}<{\centering}|m{0.8cm}<{\centering}}
\toprule
\diagbox [width=5.7em]{Scheme}{Metric} & \textbf{(A)} & \textbf{(B)} & \textbf{(C)} & \textbf{(D)} & \textbf{(E)}\\
\midrule[1pt]
Uchida's & 0.1\% & \textbf{$\geq$1,000} & 94.1\% & 21ms & 0.1\% \\
\texttt{Rand} & 0.0\% & 111 & 30.2\% & 312ms & 0.3\% \\
\texttt{WF} & 0.0\% &194 & 41.3\% & 320ms & \textbf{0.0\%}\\
\texttt{ATGF} & 0.0\% & 117 & \textbf{94.3\%} & 303ms & \textbf{0.0\%}\\
\texttt{M-S} & \textbf{0.7\%} & \textbf{$\geq$1,000} & 79.5\% & \textbf{750ms} & \textbf{0.0\%}\\
\bottomrule
\end{tabular}
\caption{Evaluation of extra security requirements w.r.t. MNIST.}
\label{table:m}
\end{table}

\begin{table}[htbp]
\centering
\begin{tabular}{m{1.6cm}<{\centering}|m{0.8cm}<{\centering}|m{1.1cm}<{\centering}|m{0.8cm}<{\centering}|m{0.8cm}<{\centering}|m{0.8cm}<{\centering}}
\toprule
\diagbox [width=5.7em]{Scheme}{Metric} & \textbf{(A)} & \textbf{(B)} & \textbf{(C)} & \textbf{(D)} & \textbf{(E)}\\
\midrule[1pt]
Uchida's & 0.2\% & \textbf{$\geq$1,000} & \textbf{95.3\%} & 20ms & \textbf{0.0\%} \\
\texttt{Rand} & 0.1\% & 312 & 41.0\% & 321ms & 1.1\% \\
\texttt{WF} & 0.1\% &473 & 36.1\% & 336ms & 1.3\%\\
\texttt{ATGF} & 0.2\% & 300 & \textbf{90.4\%} & 300ms & 1.1\%\\
\texttt{M-S} & \textbf{4.5\%} & \textbf{$\geq$1,000} & 78.0\% & \textbf{798ms} & 0.3\%\\
\bottomrule
\end{tabular}
\caption{Evaluation of extra security requirements w.r.t. CIFAR10.}
\label{table:c10}
\end{table}

\begin{table}[htbp]
\centering
\begin{tabular}{m{1.6cm}<{\centering}|m{0.8cm}<{\centering}|m{1.1cm}<{\centering}|m{0.8cm}<{\centering}|m{0.8cm}<{\centering}|m{0.8cm}<{\centering}}
\toprule
\diagbox [width=5.7em]{Scheme}{Metric} & \textbf{(A)} & \textbf{(B)} & \textbf{(C)} & \textbf{(D)} & \textbf{(E)}\\
\midrule[1pt]
Uchida's & 0.2\% & \textbf{$\geq$1,000} & \textbf{98.2\%} & 19ms & \textbf{0.0\%} \\
\texttt{Rand} & 0.9\% & 412 & 21.2\% & 458ms & 3.4\% \\
\texttt{WF} & 0.7\% &479 & 12.9\% & 433ms & 5.6\%\\
\texttt{ATGF} & 1.1\% & 410 & 90.4\% & 495ms & 4.1\%\\
\texttt{M-S} & \textbf{8.2\%} & \textbf{$\geq$1,000} & 77.5\% & \textbf{784ms} & 0.3\%\\
\bottomrule
\end{tabular}
\caption{Evaluation of extra security requirements w.r.t. CIFAR100.}
\label{table:c100}
\end{table}

\subsection{Discussions}
We observed that $\texttt{M-S}$ is optimal regarding \textbf{(A)}, since spoiling this watermark would result in the largest decrease in the DNN's normal performance.
For \textbf{(B)}, it is found that white-box schemes significantly outperformed backdoor-based ones.
As for \textbf{(C)}, only one backdoor-based scheme, $\texttt{ATGF}$, had a comparable performance as white-box schemes. 
The embedding time \textbf{(D)} for $\texttt{M-S}$ is the longest, followed by that for backdoor-based schemes. 
Uchida's is the easiest scheme regarding overwriting. 
All schemes had little impact on the convergence of the DNN model in FL \textbf{(E)}.
Although Uchida's and $\texttt{M-S}$ have met all the requirements of the decentralized OV protocol and $\texttt{Merkle-Sign}$, they are white-box schemes and the corresponding communication traffic is high.

Requirements from different protocols are sometimes contradictive against each other, e.g., the watermark capacity \textbf{(B)}.
The first two protocols require the influence of the watermark to be as small as possible so many watermarks can be embedded into the DNN, in this case a large watermark capacity is desirable.
While in DNN IP transferring, it is preferable that the watermark exerts large impact to the model so a clean model and a watermarked model can be accurately differentiated, so the watermark capacity is better to be small. 

\section{Conclusion}
To explore the applicability of IPR for deep learning models by watermarking as a service, this paper studies three scenarios and presents their respective protocols.
Our analysis shows that these protocols demand extra properties other than those discussed in designing ordinary watermarking schemes, among with some are even against each other.
Moreover, empirical studies show that current watermarking schemes cannot meet all requirements in practical protocols.
Therefore, it is necessary to formulate protocols for more real-world scenarios as well as to design watermarking schemes that meet new security properties.

\section*{Acknowledgements}
This work presented in this paper was supported by National Natural Science Foundation of China (61771310). 

\bibliographystyle{named}
\bibliography{WM.bib}

\end{document}